\documentstyle[prb,aps,epsfig,multicol]{revtex}
\begin{document}
\draft
\sloppy
%%%%%%%%%%%%%%%%%%%%%%%%%%%%%%%%%%%%%%%%%%%%%%%%
\title
{Lattice bosons in a quasi-disordered environment}
\author{R. Ramakumar$^{1}$ and A. N. Das$^{2}$}
\address{$^{1}$Department of Physics and Astrophysics, University of Delhi,
Delhi-110007, India} 
\address{$^{2}$Saha Institute of Nuclear Physics,
1/AF Bidhannagar, Kolkata-700064, India} 
\date{4 May 2013}
\maketitle
%%%%%%%%%%%%%%%%%%%%%%%%%%%%%%%%%%%%%%%%%%%%%%%%%%%%%%%%%%%%%%%%%%%%%%%%%%%
\begin{abstract}
In this paper, we study non-interacting bosons in a quasi-disordered
one-dimensional optical lattice in a harmonic potential. We
consider the case of deterministic quasi-disorder produced by an
Aubry-Andr\'{e} potential. Using exact diagonalization, we investigate
both the zero temperature and the finite temperature properties.
We investigate the localization properties by using an entanglement measure.
We find that the extreme sensitivity of the localization properties to
the number of lattice sites in finite size closed chains disappear in open chains.
This feature continues to be present in the presence of a
harmonic confining potential. The quasi-disorder is found to strongly
reduce the Bose-Einstein condensation temperature and the condensate fraction
in open chains. The low temperature thermal depletion rate of the condensate 
fraction increases
considerably with increasing quasi-disorder strength.
We also find that the critical quasi-disorder
strength required for localization increases with increasing strength
of the harmonic potential. Further, we find that the low temperature
condensate fraction undergoes a sharp drop to 0.5 in the localization transition
region. The temperature dependence of the specific heat is found
to be only marginally affected by the quasi-disorder.
\end{abstract}
\maketitle
%%%%%%%%%%%%%%%%%%%%%%%%%%%%%%%%%%%%%%%%%%%%%%%%%%%%%%%%%%%%%%%%%%%%%%%
\section{Introduction}
\label{sec1}
%%%%%%%%%%%%%%%%%%%%%%%%%%%%%%%%%%%%%%%%%%%%%%%%%%%%%%%%%%%%%%%%%%%%%%%
The prediction of the localization of Bloch waves moving through a randomly 
disordered crystal is one of the fundamental results in quantum mechanics
\cite{anderson}. 
Anderson localization of matter waves is a strong function of the 
system dimensionality. In two and lower dimensions, all the 
single particle states are localized
by non-zero disorder\cite{mott,abrahams}. Anderson localization of matter waves 
was directly observed in several experiments recently\cite{billy,roati,kondov,jendrzejewski}. 
The localization of all the single particles states in one and two dimensions
occurs for random disorder (see also the note in Ref. \onlinecite{pezzepalencia}). 
However, if the disorder distribution is
deterministic, localization occurs only beyond a critical
disorder strength. A particularly simple model in which such a transition from
the extended to the localized states occurs is the one dimensional 
Aubry-Andr\'{e} (AA) model\cite{aubryandre}. 
Among the experiments mentioned above, an experiment of particular interest for
the purpose of this paper are the studies\cite{roati}, by the LENS group,
of the localization of a Bose-Einstein condensate of non-interacting lattice bosons 
subjected to the Aubry-Andr\'{e} potential.
Our primary motivation comes from 
these studies in which a finite one-dimensional (1d) Aubry-Andr\'{e} model was
experimentally realized. There have been extensive theoretical studies
of the AA model in the past\cite{azbel,suslov,thouless,kohmoto,kohmoto2,ingold,aulbach,boers}, 
and in the recent years\cite{roux,deng,modugno,larcher,adhikari,deng2,cestari,albert,larcher2,biddle}
in response to the experimental developments mentioned above
(for recent reviews see Refs. \onlinecite{modugno2,spalencia,shapiro}).
\par
 In this paper, our main focus is the finite temperature properties of bosons 
in finite 1d optical lattices with AA quasi-disorder
in harmonic confining potentials. 
%The incommensurability parameter we consider is $q$ = ($\sqrt{5}$+1)/2. 
Before going to the studies of a Bose condensate in the
AA potential in a harmonic trap, we look at the sensitivity of localization 
properties of a single boson in finite optical lattices with open and
closed boundary conditions. We note that the open boundary conditions
%are more appropriate for the experiments involving bosons in optical lattices\cite{amico}.
are more appropriate for the experiments involving bosons in optical lattices.
We will show that the extreme sensitivity of the localization properties of
a particle in the AA model with closed boundary conditions disappear when
open boundary conditions are used. This feature is important when we consider
the many-boson system in finite lattices since now one may use any lattice size.
Though this continues to be
the overall feature after the application of an additional harmonic confining 
potential, we find a small
delocalization tendency before the complete localization transition occurs. 
This feature results from a competition between the influences of the 
harmonic potential and the quasi-disorder potential. 
Further, we go on to study Bose condensation in the AA potential. We
study the effects of increasing quasi-disorder on the condensation temperature, 
condensate fraction, and the specific heat of the system. The quasi-disorder is
found to have significant effects on some of these properties. The quasi-disorder
strongly reduces the condensation temperature and the
condensate fraction near the localization transition. 
We also find that the fall in the
condensate fraction shifts to larger quasi-disorder strengths
with increasing harmonic potential strength. Since the focus of this paper
is the effects of deterministic quasi-disorder on lattice bosons in harmonic
potentials, it is fitting for us to place these studies in
the general context of existing work on bosons in disordered
environments. In the past, the effects of random disorder on interacting
3d continuum bosons have been extensively 
investigated\cite{haung,giorgini,lopatin,astrakharchik,kobayashi,yukalov,yukalov2,falco,pilati}. 
It has been found that both the condensation temperature
and the condensate fraction decreases with increasing disorder strength.
In the case of non-interacting lattice bosons,
the condensation temperature was found\cite{dellanna,dellanna2} to decrease 
for small filling while showing
the opposite trend for large filling. There has been extensive 
studies\cite{note,fisher,soyler,lin,pisarski,kruger} on the effects of 
disorder on interacting lattice bosons within the frame-work of the 
disordered Bose-Hubbard
model. Recent experimental studies\cite{white} of interacting disordered lattice
bosons in a harmonic trap find decreasing condensate fraction with increasing 
disorder strength.
\par
The rest of this paper is organized as follows. The study of the
single particle localization properties in the open and the closed finite length
chains is given in the next section (Sec. II). The study on the influence of the 
quasi-disorder 
strength on the condensation temperature, condensate fraction, and
the specific heat is presented in Sec. III. The conclusions are given in Sec. IV.
%%%%%%%%%%%%%%%%%%%%%%%%%%%%%%%%%%%%%%%%%%%%%%%%%%%%%%%%%%%%%%%%%%%%%%%%
\section{The effects of boundary conditions on localization}
\label{sect2}
%%%%%%%%%%%%%%%%%%%%%%%%%%%%%%%%%%%%%%%%%%%%%%%%%%%%%%%%%%%%%%%%%%%%%%%
%F-1
\begin{figure}
\begin{center}
\includegraphics[angle=0.0,width=4in,totalheight=4in]{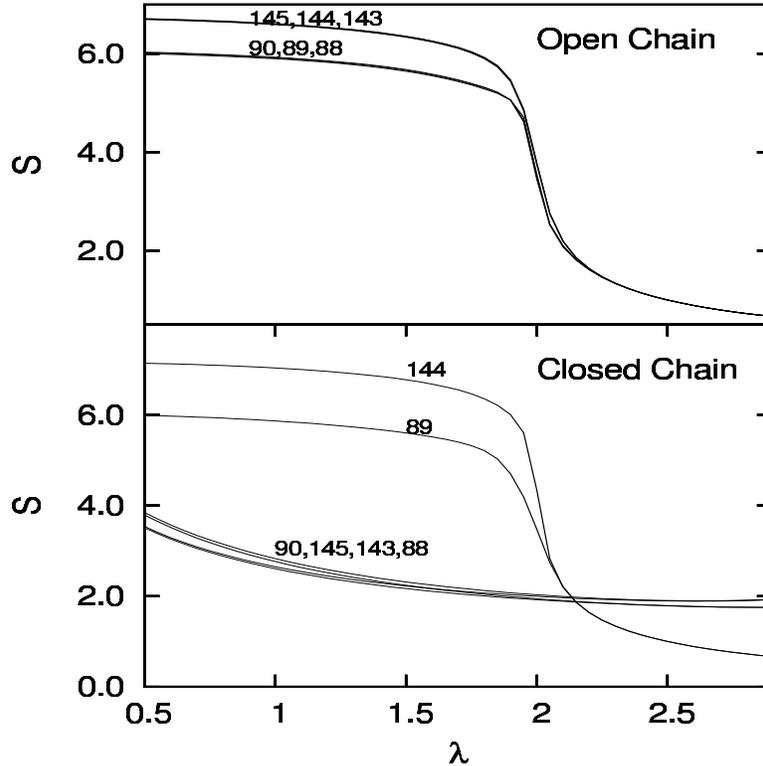}
\caption{  Entanglement (S) as a function of the AA quasi-disorder potential
strength ($\lambda$) for open and closed chains. Numbers of lattice sites in 
each chain are shown on the curves. 
}
\label{fig1}
\end{center}
\end{figure}
As stated, we consider harmonically confined lattice bosons with AA disorder
potential. The Hamiltonian of this system is: 
\begin{equation}
H=-t \sum_{<ij>}\left(c^{\dag}_{i}c_{j}+c^{\dag}_{j}c_{i}\right)
  +\sum_{i}\left[Ka^{2}i^{2}+\lambda Cos(2\pi qi)-\mu \right]c^{\dag}_{i}c_{i},
\end{equation}
where $t$ is the kinetic energy gain when a boson hop
from site $i$ to its nearest neighbor site $j$ in a one-dimensional optical lattice
with a lattice constant $a$,
$c^{\dag}_{i}$ is a boson creation operator,
$K$ the strength of the harmonic potential, $\lambda$ the
strength of the AA potential, and 
$q$ = ($\sqrt{5}$+1)/2 is the incommensurability parameter.
Here $t$, $Ka^{2}$, and  $\lambda$ have energy units.
All the energies in this paper are measured in units of $t$.
After writing this Hamiltonian in 
a single particle site basis ($\left|i\right>$), 
we numerically diagonalize it to obtain its eigen-energies and 
eigen-functions for various lattice sizes, quasi-disorder strengths,
and the harmonic potential strengths. The localization properties
of the eigen-functions are monitored by calculating the Shannon
entropy which measure the quantum entanglement\cite{stephan}. 
\newpage
The Shannon entropy is given by
\begin{equation}
S\,=\,-\sum_{i}p_{i}$ log$_{2}\,p_{i},
\end{equation}
where, 
\begin{equation}
p_{i}=\,|a_{i}|^{2} 
\end{equation}
in which the $a_i$ is the $i$th site amplitude of the ground state wave-function 
\begin{equation}
\left|\psi_G\right>$ = $\sum_{i}a_{i}\left|i\right>.
\end{equation}
The quantum entanglement is maximum for a fully extended state
and is zero for a state localized on a site.
\par
The results of 
our calculations for open and closed chains\cite{chain}, in the absence of the
harmonic potential, are shown in Fig. 1. 
In the (finite size) closed
chain results shown, the entanglement ($S$) falls abruptly at
$\lambda \approx$ 2 signifying a transition from an extended state to a 
localized state for lattices with 144 and 89 sites, but not for 
other numbers of lattice sites as shown in Fig. 1.
These numbers (89 and 144) are members of the Fibonacci series,
and the ratios between two consecutive Fibonacci numbers approaches 
$q$ = ($\sqrt{5}$+1)/2, the golden ratio, in the AA potential in 
the large system limit.
As has been noted earlier\cite{kohmoto,ingold,cestari},
the transition occurs for these numbers because of
matching of the periodic boundary conditions on
the lattice and the quasi periodicity of the AA potential. 
This extreme sensitivity disappears in the chains with open boundary
conditions, as shown in the top panel of the Fig. 1. The effect of the
harmonic potential on the localization for various values of $k = Ka^{2}$ 
is shown in Fig. 2. It is 
clear that $S$ continues to maintain the overall feature of the
$k\, =\, 0$ case. However, we find a small increase in $S$
before the complete localization 
transition occurs. This can be understood by examining 
the nature of the wave-functions shown in Fig. 3. We 
have  plotted  the  square of
%F-2
\begin{figure}
\begin{center}
\includegraphics[angle=270.0,width=3.75in,totalheight=3.75in]{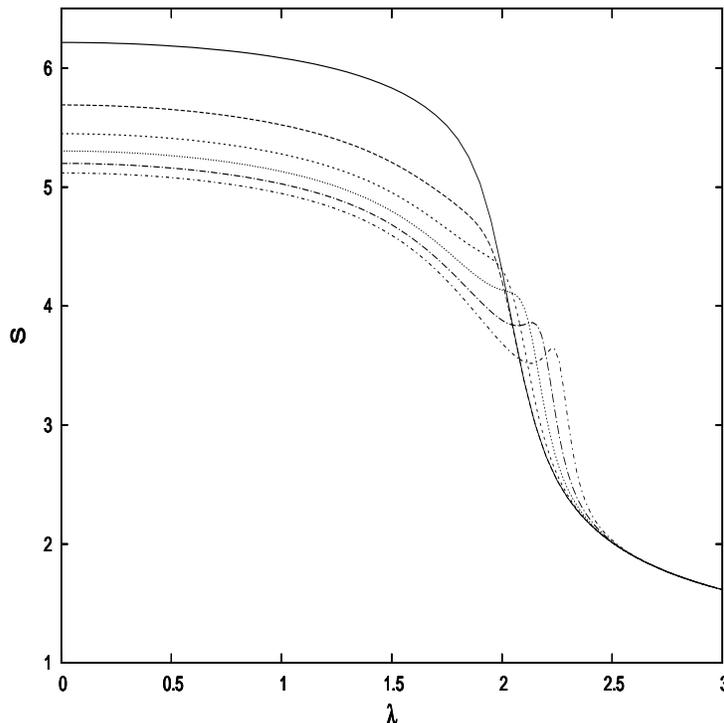}
\caption[]
{
Entanglement (S) as a function of the AA quasi-disorder potential strength ($\lambda$)
for an open chain with 100 lattice sites in harmonic potentials of different
strengths. The curves shown are for : k = 0 (solid), 0.00001 (long dash),
0.00002 (short dash), 0.00003 (dot), 0.00004 (long dash-dot), and
0.00005 (short dash-dot). The $k$ is given in units of $t$.
 }
\label{fig2}
\end{center}
\end{figure}
%F-3
\begin{figure}
\begin{center}
\includegraphics[angle=270.0,width=4in,totalheight=3.250in]{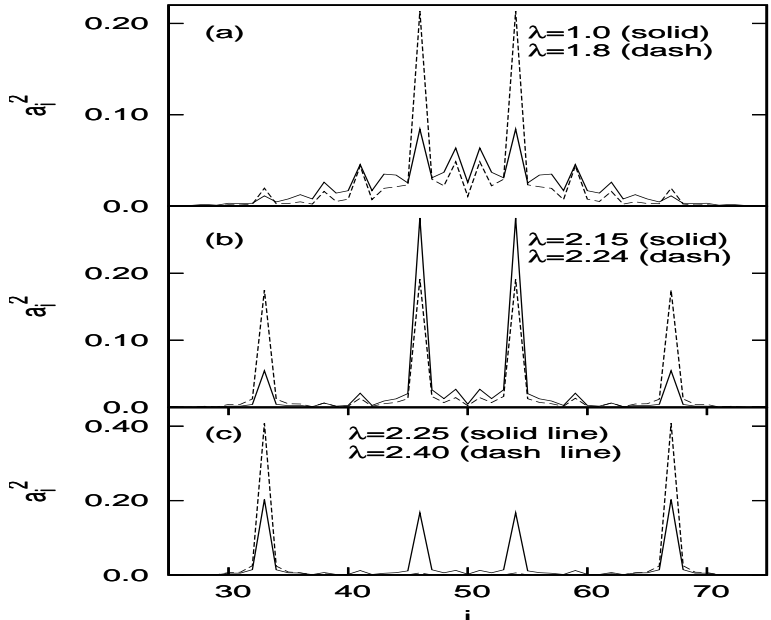}
\caption[]
{The wave functions for k = 0.00005.
}
\label{fig3}
\end{center}
\end{figure}
\noindent the amplitude of the
ground state wave function as a function of $i$ for $100$ 
sites with $k\,=\,0.00005$ and for different values of $\lambda \ge 1$. For 
this case, the $a_{i}^{2}$ exhibit two peaks around
$i\,=\,46$ and $54$. The heights of these two peaks increase
with increasing $\lambda$ up to $\lambda \,=\,2.15$. The entanglement,
correspondingly, decreases up to $\lambda \,=\,2.15$ for $k\,=\,0.00005$. 
As $\lambda$ increases further, the peak heights at $i\,=\,46$ and $54$
positions decrease, while another two peaks appear around
$i\,=\,33$ and $67$. For $\lambda$ around $2.24$ the peak
heights at these four positions become almost equal.
On further increase in $\lambda$, the peak heights at $i\,=\,46$ and $54$
decrease while those at $i\,=\,33$ and $67$ increase rapidly.
We find that the $S$ shows a maximum around $2.24$ and then decreases
rapidly. It is clear that the small increase in $S$ results
from a competition between the tendency of the
quasi-disorder potential to localize the wave function on two sets
of lattice sites away from the center and tendency of the
harmonic potential to bring the particle to the center of the trap.
This section dealt with the single boson properties. In the next
section we consider the many-boson system.
%%%%%%%%%%%%%%%%%%%%%%%%%%%%%%%%%%%%%%%%%%%%%%%%%%%%%%%%%%%%%%%%%%%%
\section{Bose condensation in the AA potential} 
\label{sec3}
%%%%%%%%%%%%%%%%%%%%%%%%%%%%%%%%%%%%%%%%%%%%%%%%%%%%%%%%%%%%%%%%%%%%
In this section we study the Bose condensation temperature, the 
temperature dependence of the condensate fraction and the specific
heat of harmonically confined lattice bosons in AA potentials\cite{1dconden}. 
%It should be clearly stated that in 1d lattices there is no phase transition.
%However, at sufficiently low temperatures macroscopic number of bosons 
%will occupy the ground state and a quasi condensate 
%will arise \cite{ketterledruten,ramdas}. 
To study these properties for a finite system,  
we first numerically diagonalize the Hamiltonian matrix to
obtain the single boson energy levels $E_i$.
In terms of $E_i$, the total number of bosons is given
by the number equation
\begin{equation}
N\,=\,\sum_{i=0}^{m}N(E_{i}),
\end{equation}
where $E_0$ and $E_m$ are the lowest and the highest
single boson energy levels, and
\begin{equation}
N(E_{i})=\frac{1}{e^{\beta\,(E_{i}-\mu)}-1} 
\end{equation}
in which $\beta\,=\,1/k_{B}T$ with $k_{B}$ the Boltzmann
constant and $T$ the temperature. 
The chemical potential and boson populations
in the various energy levels are calculated using
the boson number equation.
The specific heat is calculated
from the temperature derivative of total energy 
\begin{equation}
 E_{tot}=\sum_{i=0}^{m}N(E_{i})E_{i} 
\end{equation}
and the condensation temperature\cite{ketterledruten,ramdas}($T_0$) is
determined by solving the number equation after setting $N_{0}=0$ and 
$\mu =E_{0}$. 
The isothermal compressibility ($\kappa_T$) of non-interacting bosons 
is given by\cite{kramer} 
\begin{eqnarray}
 \kappa_T &=& \frac{1}{N} \frac{\delta N}{\delta \mu}  \\
  &=& \beta \, \left[1 + \frac{\sum_{i=0}^{m}(N(E_{i}))^2} 
{\sum_{i=0}^{m}N(E_{i})}\right].
\end{eqnarray}
We have calculated the 
isothermal compressibility for our system using Eq. (9) and studied 
its variation with temperature for different values of the AA potential
strength ($\lambda$). 
  
The effect of increasing quasi-disorder strength on the condensation 
temperature is shown in Fig. 4. The 
%F-4
\begin{figure}
\begin{center}
\includegraphics[angle=270.0,width=4in,totalheight=3.0in]{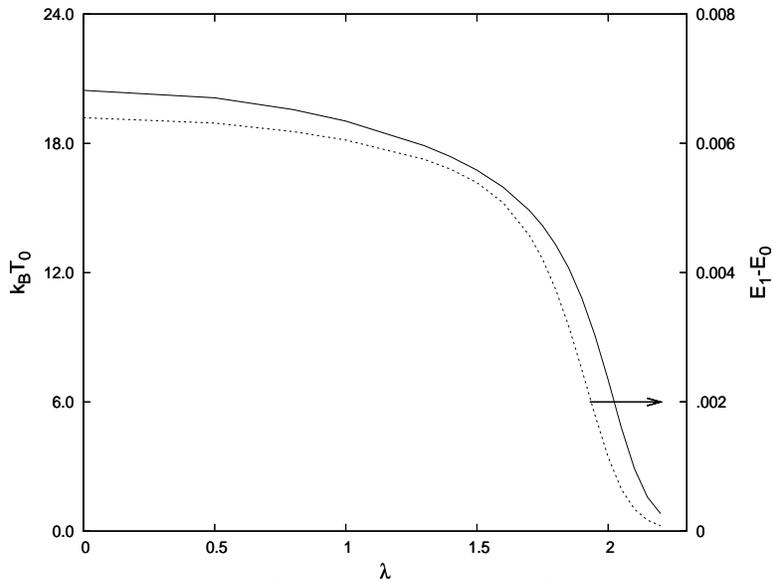}
\caption[]
{
The condensation temperature (T$_0$) as a function of $\lambda$ (solid line)
for 10000 bosons in a 1d lattice of 100 sites. The harmonic trap
strength  k = 0.00001. The dashed line shows the variation of
the difference between the energies of the first excited state (E$_1$)
and the ground state (E$_0$).
}
\label{fig4}
\end{center}
\end{figure}
%F-5
\begin{figure}
\begin{center}
\includegraphics[angle=270.0,width=4in,totalheight=4in]{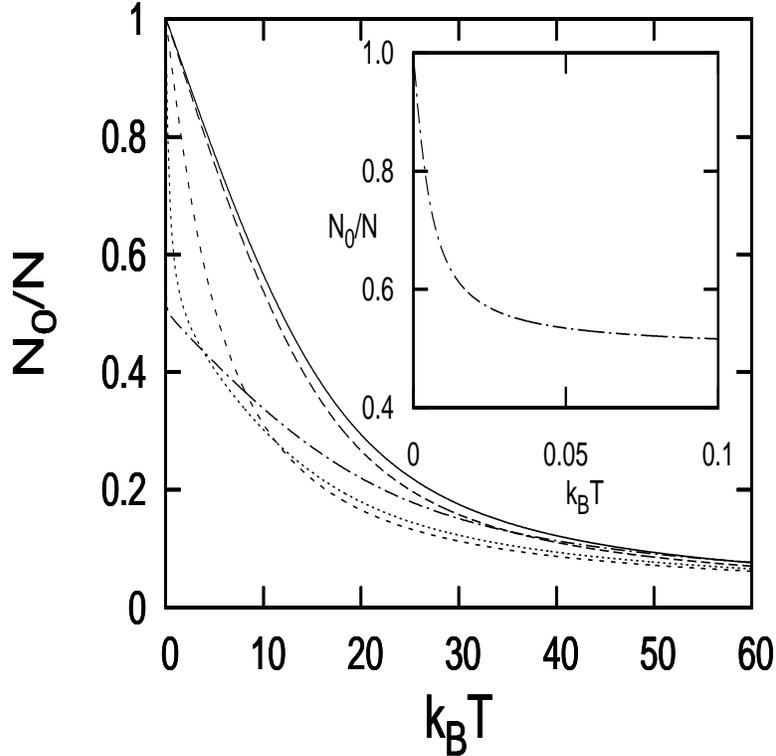}
\caption[]
{The temperature dependence of the condensate fraction for different
values of $\lambda$ for 10000 bosons in
a lattice of 100 sites in a harmonic potential of strength k = 0.00001.
The results shown are for : $\lambda$ = 0 (solid line), 1.0 (long dash),
2.0 (short dash), 2.2 (dot), and 2.5 (dash-dot). The inset shows
the low temperature variation for $\lambda$ = 2.5.
}
\label{fig5}
\end{center}
\end{figure}
\noindent quasi-disorder clearly leads to
a suppression of $T_0$. The maximum reduction occurs in the
transition region between the extended and the localized states.
We also find an interesting connection between the variation
of the $T_0$ and the quasi-disorder dependence of the energy difference
($E_1 - E_0$) between the first excited state and the ground 
state. The $T_0$ is found to closely track ($E_1 - E_0$).
The reduction of ($E_1 - E_0$) leads to an increase in the
one-boson density of energy states at the bottom of the
energy spectrum. This leads to the reduction in $T_0$. 
The temperature  dependence 
of the ground state
occupancy with various quasi-disorder strengths is shown in Fig. 5.
With increasing quasi-disorder strength, the thermal depletion rate
is found to increase considerably at low temperatures. This is especially
significant as one moves in to the localization transition region.
In the localized limit, the condensate fraction decreases sharply with a slight
increase  in  temperature  as is clear from the results for $\lambda \, =\,2.5$.
%F-6
\begin{figure}
\begin{center}
\includegraphics[angle=270.0,width=4in,totalheight=3in]{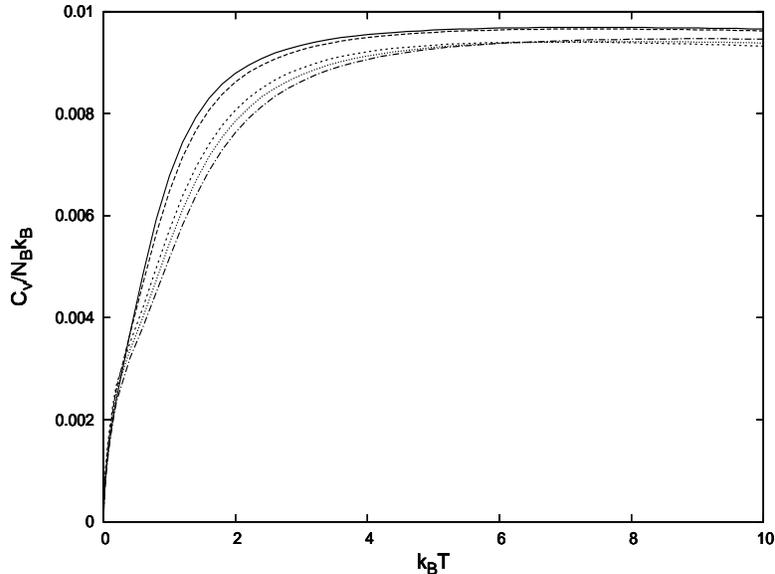}
\caption[]
{
The temperature dependence of the specific heat for different
values of $\lambda$ for 10000 bosons in
a lattice of 100 sites in a harmonic potential of strength k = 0.00001.
The results shown are for : $\lambda$ = 0 (solid line), 1.0 (long dash),
2.0 (short dash), 2.2 (dot), and 2.5 (dash-dot).
}
\label{fig6}
\end{center}
\end{figure}
The temperature dependence of the specific heat for various values
of the quasi-disorder strength is shown in Fig. 6. 
Even though in the range of quasi-disorder strengths shown, the 
system has undergone the localization transition, the specific heat 
is found to be only marginally affected by the quasi-disorder. 
%F-7
\begin{figure}
\begin{center}
\includegraphics[angle=270.0,width=4in,totalheight=3in]{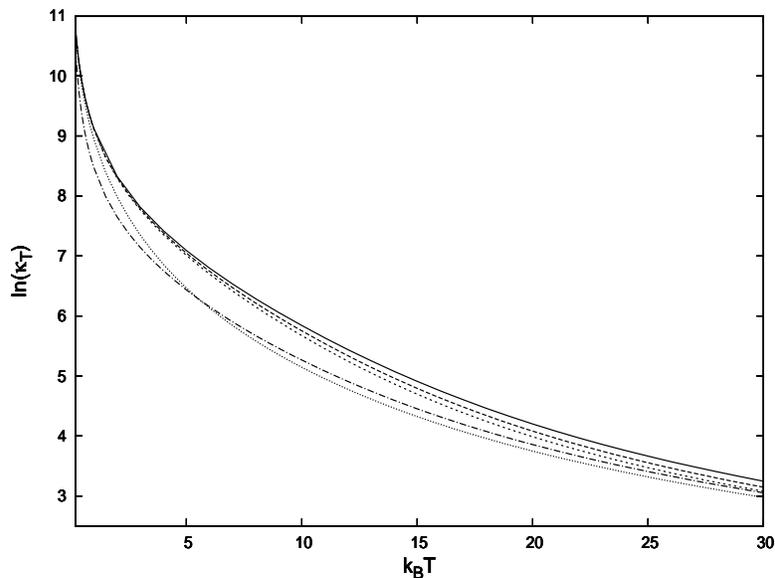}
\caption[]
{
The temperature dependence of $\kappa_T$ for different
values of $\lambda$ for 10000 bosons in
a lattice of 100 sites in a harmonic potential of strength k = 0.00001.
The results shown are for : $\lambda$ = 0 (solid line), 1.0 (long dash),
1.3 (short dash), 2.0 (dot), and 2.2 (dash-dot).
}
\label{fig7}
\end{center}
\end{figure}
In Fig. 7 we showed the variation of the 
isothermal compressibility with temperature for different $\lambda$ 
values. The compressibility increases sharply with decreasing temperature 
as expected from its expression in Eq. (9), since both the $1/T$ and 
$\sum (N(E_i))^2/N$ terms in $\kappa_T$ increase with decreasing 
temperature. Regarding $\lambda$ dependence of $\kappa_T$, we find 
that at high temperatures it decreases with increasing $\lambda$ 
up to $\lambda=2$, beyond which, as the system enters into the 
localized regime, $\kappa_T$ increases with increasing $\lambda$. 
At low temperatures $\kappa_T$ decreases with $\lambda$ even beyond
$\lambda=2$. 
This behavior is similar to that observed 
for $N_0/N$ in Fig. 5. It may be mentioned that the term 
$\sum (N(E_i))^2/N$ in $\kappa_T$ is mainly determined by its 
ground state contribution, $i.e.$ $N_0^2/N$, when $N_0$ reaches 
a macroscopic value.  
\par
The effect of increasing quasi-disorder 
strength on the thermal depletion of the
condensate fraction was shown in Fig. 5. We now go on to a detailed 
investigation of the quasi-disorder induced depletion of the condensate.
%F-8
\begin{figure}
\begin{center}
\includegraphics[angle=270.0,width=4in,totalheight=3in]{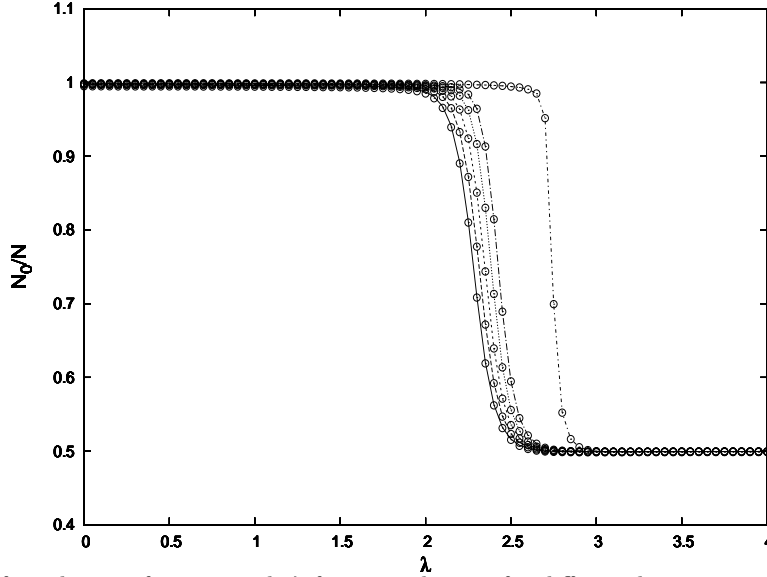}
\caption[]
{
The variation of condensate fraction with $\lambda$ for 10000 bosons
for different harmonic potential strengths.
The lines are for 100 sites for k values: 0.00001 (solid), 0.00002 
(long-dash), 0.00003 (short-dash), 0.00004 (dot), 0.00005 (long dash-dot), 
and 0.0001 (short dash-dot). The circles are for 300 sites for the 
same k values. The temperature is k$_B$T = 0.1.
}
\label{fig8}
\end{center}
\end{figure}
In Fig. 8, we present the variation of the condensate fraction with quasi-disorder
strength for different harmonic potentials. 
We find that the increase in quasi-disorder strength leads to two major effects. 
The first one is the sharp drop in
the condensate fraction to 0.5 in the localization transition region and  
second is the shift of this fall to higher values of quasi-disorder strengths
with increasing strength of the harmonic confinement. The latter result 
correlates well with the localization properties given in Fig. 2.
This shift is due to larger quasi-disorder strengths required for the localization
transition when the harmonic potential increases.
In order to understand the first effect, we
examined the energy level differences between the ground state
and the lower excited energy levels. In Fig. 9, we have shown the
energy level difference, as a function of quasi-disorder strength, 
between the first excited state and the ground state, 
and between the second excited state and the ground state.   
When the energy difference between the first excited state and the ground state
becomes very very small, both the states are
almost equally populated, and $N_{0}/N$ reduces to about 0.5.
Note that the energy difference
between the first excited and the ground state remains finite, though small,
even for large values of $\lambda$, which is not visible in the scale of this
figure. Note also that that the $E_2-E_0$ shows a minimum in the vicinity of 
the localization transition and then increases fast as the system moves
deeper in to the localized regime. 
%F-9
\begin{figure}
\begin{center}
\includegraphics[angle=270.0,width=4in,totalheight=4in]{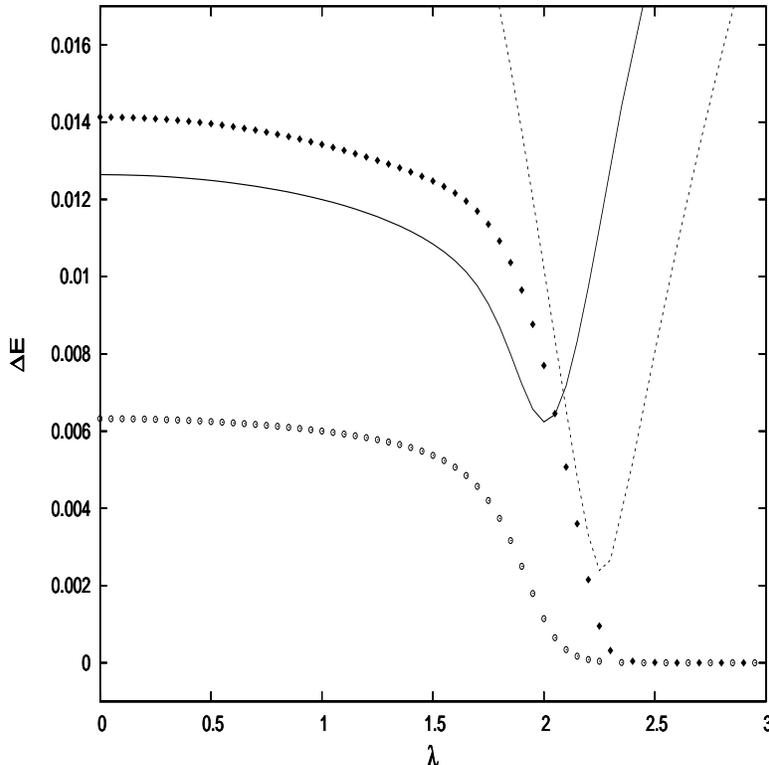}
\caption[]
{
The variation of the energy difference $\Delta$E between low lying levels
with $\lambda$. The results are for k = 0.00001 : $E_1-E_0$ (open circles)
and $E_2-E_0$ (solid line) and for k = 0.00005 : $E_1-E_0$ (filled diamonds)
and $E_2-E_0$ (dash line).
}
\label{fig9}
\end{center}
\end{figure}
%%%%%%%%%%%%%%%%%%%%%%%%%%%%%%%%%%%%%%%%%%%%%%%%%%%%%%%%%%%%%%%%%%%
\section{Conclusions}
\label{sec4}
%%%%%%%%%%%%%%%%%%%%%%%%%%%%%%%%%%%%%%%%%%%%%%%%%%%%%%%%%%%%%%%%%%%
In this paper, we investigated some properties of a single boson
as well as many boson systems in finite one-dimensional optical lattices
with Aubry-Andre quasi-disorder in harmonic confinements. We find that
the open lattices do not show the extreme sensitivity of the 
localization properties seen in finite closed lattices. This continues to be
the case in the presence of an overall harmonic confining potential.
In the presence of the harmonic potential, we find a small delocalization
tendency before a completely localized state is reached. On the basis of
the quasi-disorder induced changes occurring in the single particle wave-function,
we argued that this tendency is due to a competition between the effects
of the harmonic potential and the quasi-disorder potential. For the many-boson
system, we studied the effects of quasi-disorder on the condensation temperature,
the condensate fraction, and the specific heat. We find that 
the condensation temperature decreases with increasing quasi-disorder strength,
the rate of the reduction is maximum in the localization transition region.
The quasi-disorder is found to have a significant effect on the 
condensate fraction. With increasing quasi-disorder, the thermal depletion
rate is found to increase considerably, especially in the localization
transition region. In the localized limit, the condensate fraction
decreases significantly with a small increase in the temperature. 
We also studied the variation of the low temperature condensate
fraction with increasing quasi-disorder strength. We find that
the condensate fraction shows a sharp drop to 0.5 in the localization
transition region. We also find that this fall shifts to higher
quasi-disorder strengths with increasing strength of the harmonic
confining potential. The effect of quasi-disorder on the specific 
heat is found to be a marginal reduction in a certain 
range of temperature.  
%%%%%%%%%%%%%%%%%%%%%%%%%%%%%%%%%%%%%%%%%%%%%%%%%%%%%%%%%%%%%%%%%
\acknowledgments
RRK thanks Professor Milan Sanyal, Director, SINP and
Professor Bikas Chakrabarti, Head, TCMP Division, SINP
for hospitality at SINP. RRK also thanks 
Professor Laurent Sanchez-Palencia (LCF, Palaiseau) for drawing
his attention to two recent papers from his group.
%%%%%%%%%%%%%%%%%%%%%%%%%%%%%%%%%%%%%%%%%%%%%%%%%%%%%%%%%%%%%%%%%%

%%%%%%%%%%%%%%%%%%%%%%%%%%%%%%%%%%%%%%%%%%%%%%%%%%%%%%%%%%%%%%%%%%%%%%

\begin{references}
\bibitem{note}
There has been intensive studies of the disordered Bose-Hubbard model since the 
late eighties (Ref. \onlinecite{fisher}).
Since we are not considering interacting bosons in this paper, we request the reader
to consult the recent works (Refs. \onlinecite{soyler,lin,pisarski,kruger}), and the
references contained therein.
\bibitem{chain}
In the numerical calculations, the open-chain and the closed chain results are
obtained by taking $t_{N_{s},1}=t_{1,N_{s}}=0$ and 
$t_{N_{s},1}=t_{1,N_{s}}=t$, respectively, where $N_{s}$ is the number of sites.
\bibitem{1dconden}
Since our system is one-dimensional, one has a quasi-condensate at finite temperatures.
\end{references}
\end{document}